\pdfoutput=1 
\documentclass[prr,twocolumn,longbibliography,preprintnumbers,superscriptaddress,showpacs]{revtex4-1}

\usepackage{graphicx}
\usepackage{bm}
\usepackage{amsmath}
\usepackage{amssymb}
\usepackage{slashed}
\usepackage[colorlinks,pdfstartview=FitH]{hyperref}
\hypersetup{linkcolor=blue,citecolor=blue,filecolor=black,urlcolor=blue}

\def \be {\begin{equation} }
\def \ee {\end{equation}}

\def \bem {\begin{multline}}
\def \eem {\end{multline}}

\def \bes {\begin{subequations} }
\def \ees {\end{subequations}}

\def \pd {\partial}

\def \b {\beta}
\def \c {\chi}
\def \d {\delta}
\def \e {\epsilon}
\def \ce {\varepsilon}
\def \g {\gamma}
\def \k {\kappa}

\def \p {\pi}

\def \l {\lambda}

\def \n {\nu}

\def \t {\tau}

\def \x {\xi}
\def \z {\zeta}
\def \f {\phi}
\def \vf {\varphi}
\def \bvf {{\bar \varphi}}

\def \D {\Delta}
\def \G {\Gamma}

\def \L {\Lambda}

\def \<{\langle}
\def \>{\rangle}
\def \+{\dagger}
\def \({\left(}
\def \){\right)}
\def \[{\left[}
\def \]{\right]}

\def \kz {\text{KZ}}
\def\eff {\text{eff}}

\usepackage[normalem]{ulem}

\begin{document}

\author{Chang~Lei}
\affiliation{School of Physics and Astronomy, Sun Yat-Sen University, Zhuhai 519082, China}
\author{Shu~Lin}
\affiliation{School of Physics and Astronomy, Sun Yat-Sen University, Zhuhai 519082, China}
\affiliation{Guangdong Provincial Key Laboratory of Quantum Metrology and Sensing, Sun Yat-Sen University, Zhuhai 519082, China}

\title{Phases of Quench Dynamics in the Presence of Fluctuation}
\date{\today}

\begin{abstract}
We study the effect of thermal fluctuations on a sourced quench in a system with $Z_2$ symmetry. By ignoring the fluctuation of finite momentum modes and tracing the dynamics of zero momentum mode driven by a spatially homogeneous source near the critical point, we map out a phase diagram for the quench dynamics. The phase diagram consists of three different phases. Phase I occurs for large fluctuations with the relaxation time set by fluctuation induced inverse effective mass square. Phase II occurs for small fluctuation and slow quench rate. The dynamics is characterized by a modified Kibble-Zurek scaling. Phase III corresponds to small fluctuations and rapid quench rate. The relaxation time tends to a finite value in the rapid quench limit. We also estimate the fluctuations of finite momentum modes, finding significant enhancement of the effective mass square. We speculate the qualitative features of the phase diagram may remain the same, but enhanced fluctuations can lead to shrinkage of phase II and III.
\end{abstract}
\maketitle 

\section{\bf Introduction}%

Dynamics near a critical point has attracted significant interests over the past decades. A key feature is the divergence of relaxation time as the system approaches the critical point. It has been shown in seminal works by Kibble and Zurek (KZ) that the system reaches a non-adiabatic state when it evolves close enough to the critical point. At the breaking down of adiabaticity the system is frozen in a state, exhibiting universal scaling behaviors, termed KZ mechanism \cite{Kibble:1976sj,Kibble:1980mv,Zurek:1985qw}. The mechanism has broad impacts in a variety of fields such as condensed matter physics \cite{Chuang:1991zz,Bowick:1992rz}, cold atom physics \cite{Ruutu:1995qz,Baeuerle:1996zz} and heavy ion physics \cite{Mukherjee:2016kyu,Akamatsu:2018vjr}.

In reality, evolution of a system across the crtical point is ususally driven by variation of temperature or variation of external source field, corresponding to thermal quench and sourced quench respectively. At rapid quench rate, deviation of the KZ scaling is known from experiment \cite{Dodd:1998aan,Ko:2019ntc,Goo:2021nyg} and numerical simulation \cite{PhysRevB.89.054307,PhysRevB.90.134108,Chesler:2014gya,delCampo:2021rak} for thermal quench. Different explanations of the deviation has been proposed \cite{PhysRevLett.102.070401,PhysRevB.89.054307,PhysRevB.90.134108,Chesler:2014gya,Xia:2021xap}, yet no consensus has been reached.

\begin{figure}[htbp]
     \begin{center}
          \includegraphics[height=5cm,clip]{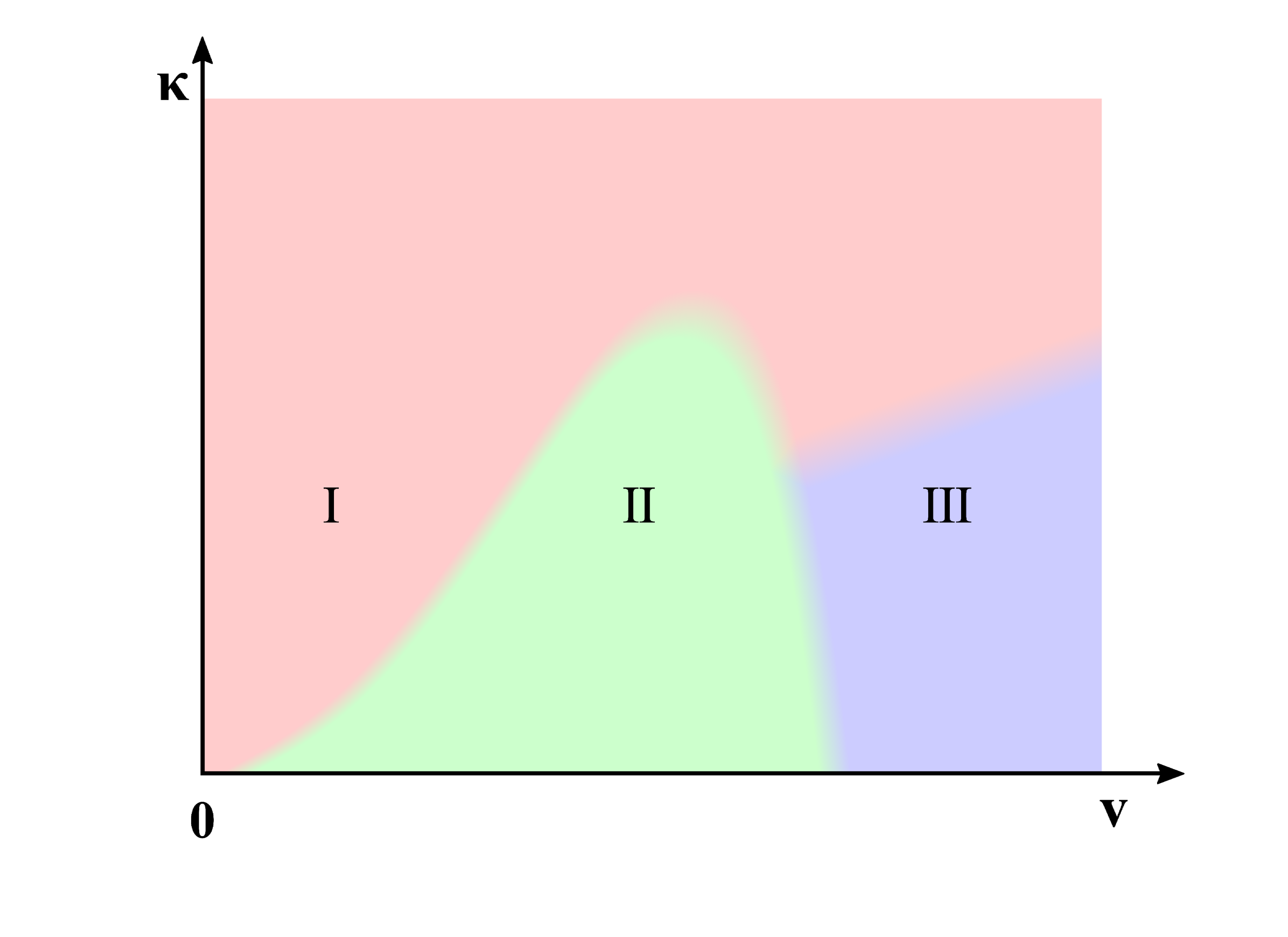}
          \caption{Schematic plot of ``phase diagram'' for the quenching dynamics. The axes $\k$ and $v$ characterize the strength of fluctuation and quench rate. Three different phases are identified using a dynamical relaxation time $t_0$ as ``order parameter'' (precise definition in the text). An approximate KZ scaling appears as phase II with $t_0\sim v^{-\#}$, with a noise dependent exponent. Phase I corresponds to a slow quench scenarios. It is featured by breaking down of mean field approximation with a different scaling $t_0\sim \k^{-1/2}$. Phase III corresponds to a rapid quench scenario, with $t_0$ tending to a $\k$-dependent constant as $v\to\infty$. There is no sharp transitions between the phases.}
    \label{fig:pd}
    \end{center}
\end{figure}

In this paper, we introduce a new ingredient thermal fluctuations into the evolution of a system across the critical point. Thermal fluctuation exist in any finite temperature system and are known to modify the long time hydrodynamic behaviors \cite{POMEAU197563}. In modern formulation of hydrodynamic effective field theory, fluctuations affect long time hydrodynamic behaviors through generating correction to dispersion of the corresponding hydrodynamic degree of freedoms \cite{Kovtun:2012rj,Kovtun:2011np,Kovtun:2014hpa,Akamatsu:2016llw,Akamatsu:2017rdu}. One may expect that fluctuations can play a similar role in dynamics near the critical point to that in hydrodynamics. We will find that it is indeed the case and fluctuations can lead to significant modification of the scaling behaviors. A schematic plot for ``phase diagram'' of quenching dynamics is shown in Fig.~\ref{fig:pd}. In particular, a modified KZ scaling appears as phase II only below a critical strength of fluctuation in a finite window of quench rate. 

The paper is organized as follows: in Section~\ref{sec_quench}, we study a sourced quench process in a stochastic model. We first obtain KZ scaling and its violation at slow and rapid quench rate respectively in the absence of fluctuations. Then we study the effect of fluctuations by both numerical and analytical methods. In Section~\ref{sec_finite}, we revisit the simplifications made in Section~\ref{sec_quench}, in particular we give an estimate of the fluctuations of finite momentum modes, finding a significant enhancement of the fluctuations for the zero momentum mode. We summarize in Section~\ref{sec_summary}.

\section{\bf Quench in a Stochastic Model}\label{sec_quench}

We start with the following stochastic equation
\begin{align}\label{eom_f}
\frac{d\f(t,x)}{dt}+\g \frac{\d \G}{\d \f(t,x)}=\c(t,x),
\end{align}
with $\f(t,x)$ being the order parameter field. $\G$ is the free energy functional assuming $Z_2$ symmetry
\begin{align}\label{Gamma}
\G=\int d^3x\(\frac{1}{2}\(\nabla\f\)^2+\frac{1}{2}m^2\f^2+\frac{1}{4}\l\f^4+J(t,x)\f(t,x)\).
\end{align}
$\g$ is a damping parameter. The field $\x(t,x)$ represents the stochastic force, which is taken to be Gaussian white noise
\begin{align}\label{white}
\<\c(t,x)\>=0,\quad \<\c(t,x)\c(t',x')\>=2\g T\d(t-t')\d^{(3)}(x-x').
\end{align}
This belongs to model A in the classification of dynamical universality class by Hohenberg and Halperin \cite{Hohenberg:1977ym}. Correlation functions of this model have been studied in \cite{PhysRevB.86.064304,Schweitzer:2020noq,Schaefer:2022bfm}. We consider a critical point of the mean field type, for which we have the following scalings for parameters $m\sim \x^{-1}$, $\l\sim \x^{0}$ close to the critical point. $\x$ is the correlation length, which diverges as
\begin{align}
\x\sim \e_T^{-\n},\quad \e_T=\frac{T-T_c}{T_c}.
\end{align}
$\e_T$ measures the deviation from the critical point and $\n$ is a critical exponent. We will be interested in dynamics close to the critical point, for which we can ignore the mass term provided that $\f$ is not parametrically as small as $\x^{-1}$. To quantify the effect of fluctuations, we consider a sourced quench process driven by a prescribed spatially homogeneous source field $J(t)$. We can set $\g=1$ in \eqref{eom_f} and \eqref{Gamma}, which amounts to a rescaling in $t$. We can further set $\l=1$ by the following rescalings $\f=\f'\l^{-1/2}$, $J=J'\l^{-1/2}$ and $\c=\c'\l^{-1/2}$. 
Dropping the primes for notational simplicity, we have the following equations
\begin{align}\label{eom_fp}
\frac{d\f(t,x)}{dt}-\nabla^2\f(t,x)+\f(t,x)^3+J(t)=\c(t,x).
\end{align}
In the absence of noise, the source $J(t)$ only excites zero momentum mode defined as $\vf(t)=\frac{\int d^3x\f(t,x)}{V}$, which satisfies
\begin{align}\label{eom_vfm}
\frac{d\vf(t)}{dt}+\vf(t)^3+J(t)=0.
\end{align}
To be specific, we consider the following source $J=\tanh(vt)$, with $v$ setting the scale for quench rate. When the quench is slow, it is well-known that the dynamics exhibits KZ scaling, which we now review.

Very close to the critical point, the effect of the quenching source can be mapped to deviation of the critical point by identifying the corresponding vevs using critical exponents \cite{Bu:2019epc}.
\begin{align}\label{mapping}
J^{1/\d}\sim \<\vf\>\sim \e_T^{\b}.
\end{align}
When $|t|\gg v^{-1}$, $J$ varies very slowly so that we can drop the time derivative term in \eqref{eom_vfm}. It follows that $\vf$ evolves adiabatically as $\vf=(-J)^{-1/3}$. When $|t|\simeq v^{-1}$, $\vf$ start to approach the critical value $\vf=0$ and adiabaticity is expected to break down due to critical slowing down. This occurs when the the ``relaxation time'' of the source $t$ becomes comparable to the relaxation time of the system $\t(t)$. Here the relaxation time scales as $\t\sim\x^{-z}\sim \e_T^{-z\n}$, with the time dependence coming from $J$ through \eqref{mapping}. This condition determines a KZ time $t_\kz$, from which on the system enters a non-adiabatic regime:
\begin{align}\label{taukz}
t_\kz\sim \t(t_\kz)\sim \x^{-z}\sim \e_T^{-z\n}\sim J^{-z\n/\b\d},
\end{align}
Near $t=0$, we may replace $J\simeq vt_\kz$ to obtain
\begin{align}\label{KZ_t}
t_\kz\sim v^{-\frac{\z}{\z+1}},
\end{align}
with $\z=z\n/\b\d$. Using the mean field exponents $\d=3$, $\b=\frac{1}{2}$, $\n=\frac{1}{2}$ and $z=2$ for model A\footnote{Small deviation of the mean field exponents is known from both analytic \cite{Zinn-Justin:2002ecy,Alday:2015ota} and numerical approaches \cite{Schweitzer:2020noq,Schaefer:2022bfm}. The deviation is not essential in our analysis to follow.}, we have $t_\kz\sim v^{-2/5}$. $\vf$ also exhibits a scaling form: 
\begin{align}\label{scaling}
\vf\sim J^{1/3}F(t/t_\kz)\sim (vt_\kz)^{1/3}F(t/t_\kz)\sim v^{1/5}F(tv^{2/5}),
\end{align}
where the factor $J^{1/3}$ comes from \eqref{mapping} valid at the boundary of the KZ regime and the function $F(t/t_\kz)$ captures the non-adiabatic dynamics.
The scaling form \eqref{scaling} can also been derived by an adiabatic expansion of the mean field equation \eqref{eom_vfm} in \cite{Das:2011nk}.

The derivation above relies on the approximation $J=\tanh(vt)\simeq vt$, i.e. the onset time of non-adiabaticity $t_\kz$ is comparable to the scale of source variation $v^{-1}$. Using $t_\kz\sim v^{-2/5}$, we can see this condition is clearly violated at $v\gg1$ for a rapid quench.
We can have some analytic insight on the violation in our model: for very rapid quench, $v\to\infty$ we may replace $J$ by $2\Theta(t)-1$. This corresponds to a sudden change of potential $\frac{\vf^4}{4}-\vf\to\frac{\vf^4}{4}+\vf$. At $t<0$, the system is simply static with $\vf=1$. At $t>0$, the system evolves in the new potential with the boundary condition $\vf(t=0)=1$ by continuity. $\vf$ can be solved in terms of inverse function as
\begin{align}\label{rapid}
&\frac{1}{18}\(\sqrt{3}\p+6\ln 2\)-t=\frac{\tan^{-1}(\frac{-1+2\vf}{\sqrt{3}})}{\sqrt{3}}+\frac{1}{3}\ln(\vf+1)\nonumber\\
&-\frac{1}{6}\ln(1-\vf+\vf^2).
\end{align}
When $\vf$ crosses zero, \eqref{rapid} gives a crossing time
\begin{align}
t_0=\frac{1}{18}\(2\sqrt{3}\p+6\ln 2\)\simeq 0.84.
\end{align}
This is a limiting value for $t_0$ determined by the shape of the new potential. It clearly differs from prediction of KZ scaling, which gives $t_\kz\to0$ as $v\to\infty$.

It is suggestive to separate the two scenarios using $t_0$ as a dynamical relaxation time, which measures the lagging time of field $\vf$ in tracing the source. The two scenarios in the absence of fluctuation appears as phase II and III on the horizontal axis of the phase diagram in Fig.~\ref{fig:pd}.


The main efforts of this paper is to study how fluctuations change the picture. We first derive the following stochastic equation for $\vf$ by averaging \eqref{eom_fp} over volume
\begin{align}\label{eom_vf}
\frac{d\vf(t)}{dt}+\frac{\int d^3x\f(t,x)^3}{V}+J(t)=\c(t).
\end{align}
$\c$ is the volume averaged noise, which satisfies
\begin{align}
\<\c(t)\>=0,\quad \<\c(t)\c(t')\>=\k\d(t-t'),
\end{align}
with $\k=\frac{2T}{V}$ being the noise strength, which depends on the temperature and volume. In the following we treat it as a free parameter. The nonlinear term in \eqref{eom_vf} contains coupling between the zero momentum mode $\vf$ and other finite momentum modes. In this section, we will truncate to contribution from $\vf$ only, i.e. we set $\frac{\int d^3x\f(t,x)^3}{V}=\vf(t)^3$. We will address the limitation of the truncation in the next section.

We then proceed to solve the stochastic equation numerically for different values of $v$ and $\k$. For larger $\k$, the solution of the stochastic equation becomes noisy, so we use a timestep set by strength of noise $\D t\sim 10^{-2}\k^{-1}$. The initial condition is set at a sufficient early time $|t_s|\gg v^{-1}$ with $\vf(t_s)=\[-J(t_s)\]^{1/3}$, which is simply the vev induced by constant source in the noise free limit. For each parameter set, we generate $10^4$ independent numerical solutions and perform statistical averages to obtain $\bvf=\<\vf\>$, $\<\D\vf^2\>=\<(\vf-\bvf)^2\>$ etc.

In the presence of fluctuation, we define $t_0$ as the time when average field $\bvf$ crosses zero. In Fig.~\ref{fig:pd}, we plot a ``phase diagram'' for the quench dynamics using $t_0$ as ``order parameter''. It contains three phases characterized by different scaling behaviors: I. $t_0\sim \k^{-1/2}$; II. $t_0\sim v^{-\#}$ with the exponent $\k$-dependent and III. $t_0$ tends to a constant as $v\to\infty$. Phase II and III are qualitatively similar to their noise free counterparts. Phase I is entirely new, featured by breaking down of mean field approximation.
The region of II shrinks with increasing $\k$. Above certain critical value of $\k$, phase I and III merge, with $t_0$ sharing the $\k$ and $v$ dependence from both phase I and III, i.e. $t_0\sim \k^{-1/2}$ and tends to a constant as $v\to\infty$. We stress that there is no sharp phase boundaries in the phase diagram. The transition from one phase to another is smooth.

To have a better understanding of different phases, we split \eqref{eom_vf} into equations for the mean $\bvf$ and fluctuation $\D\vf$ respectively
\begin{subequations}
\begin{align}
&\frac{d\bvf}{dt}+3\bvf\<\D\vf^2\>+\bvf^3+\<\D\vf^3\>+J=0,\label{mean}\\
&\frac{d\D\vf}{dt}+3\bvf^2\D\vf+3\bvf(\D\vf^2-\<\D\vf^2\>)+(\D\vf^3-\<\D\vf^3\>)=\c\label{fluc}.
\end{align}
\end{subequations}
In the presence of fluctuation, the variance $\<\D\vf^2\>$ is nonvanishing in general. It acts as an effective mass square $m_{\text{eff}}^2\sim \<\D\vf^2\>$ for $\bvf$, giving rise to a finite relaxation time $\t\sim m_{\text{eff}}^{-2}$. For sufficient slow quench, the finite relaxation always justifies adiabatic approximation, invalidating KZ mechanism. The slow quench scenario corresponds to phase I, where KZ scaling is lost! Instead, it is characterized by a different scaling we now set out to obtain.

\begin{figure}[htbp]
     \begin{center}
          \includegraphics[height=5cm,clip]{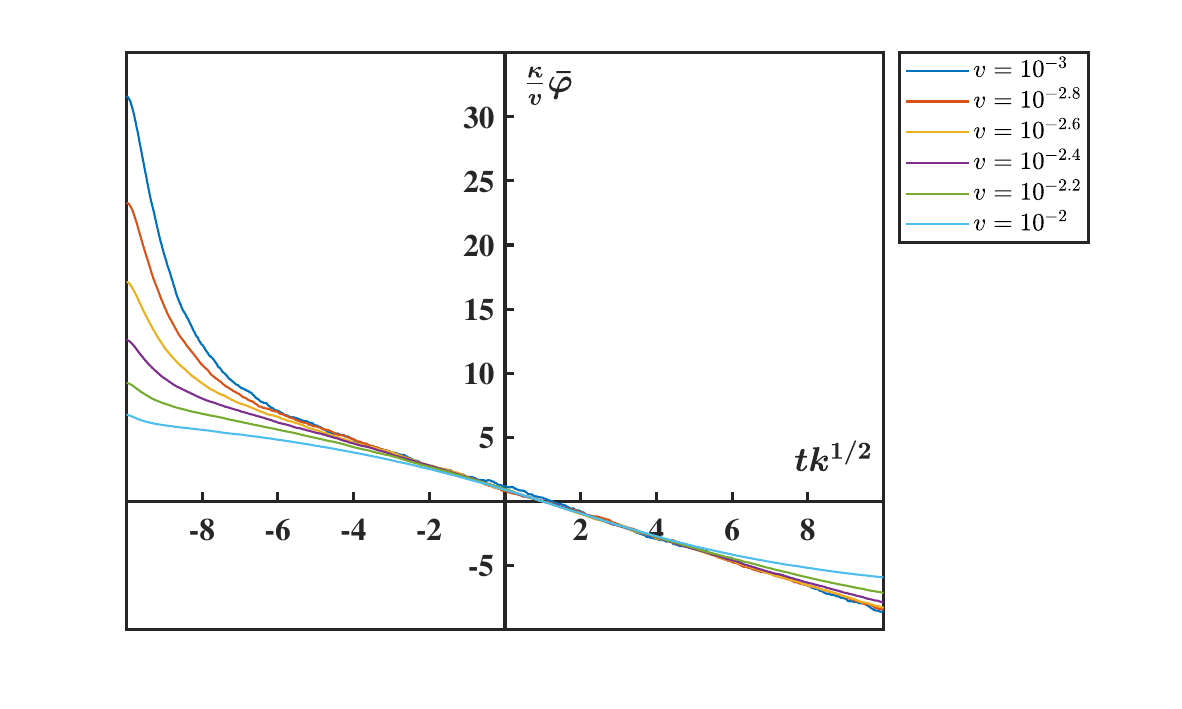}
          \caption{$\frac{\k}{v}\bvf(t)$ as a function of $t\k^{1/2}$ for $\k=0.1$ at different quenching rate $v$ in phase I. We have also extracted from independent numerical solutions that $a\simeq0.48$, $ac\simeq -0.4$, which give rise to an intercept $t_0=\frac{1}{(3+c)a\k^{1/2}}$ consistent with the plot.}
    \label{fig:I_scaling}
    \end{center}
\end{figure}

For very slow quench, the system spends significant amount of time near $\bvf\simeq0$, for which mean field approximation breaks down completely, $\<\D\vf^2\>\gg\bvf^2$. In this case we can drop the $\bvf$ dependent terms in \eqref{fluc}. The last term $\<\D\vf^3\>$ is at least linear in $\bvf$ because odd moments of $\D\vf$ vanishes when $\bvf=0$ so that this term can also be dropped. We are left with the term $\D\vf^3$, which we estimate by $\sim\D\vf\<\D\vf^2\>$. \eqref{fluc} is now reduced to a Langevin equation with $\<\D\vf^2\>$ being the effective friction. The solution of the Langevin equation gives
\begin{align}
\<\D\vf^2\>\sim\frac{\k}{\<\D\vf^2\>}\Rightarrow \<\D\vf^2\>\sim\k^{1/2}.
\end{align}
This suggests the following parametrization for $\<\D\vf^2\>$
\begin{align}\label{para2}
\<\D\vf^2\>=a\k^{1/2}.
\end{align}
As argued above, we parametrize $\<\D\vf^3\>$ by a linear dependence in $\bvf$
\begin{align}\label{para1}
\<\D\vf^3\>=c\bvf\<\D\vf^2\>.
\end{align}
We have confirmed the parametrizations \eqref{para2} and \eqref{para1} with numerical solutions to the original stochastic equation \eqref{eom_vf}. With \eqref{para2} and \eqref{para1} and $J\simeq vt$ near the origin, we find \eqref{mean} adopts the following solution
\begin{align}\label{mfb}
\bvf=A(t-t_0),
\end{align}
with
\begin{align}\label{At0}
A=-\frac{v}{(3+c)a\k^{1/2}},\quad
t_0=\frac{1}{(3+c)a\k^{1/2}}.
\end{align}
%
The form of the dynamical relaxation time $t_0$ can be easily understood from inverse of effective mass square, which is identified in \eqref{mean} as $3\<\D\vf^2\>+\frac{\<\D\vf^3\>}{\bvf}$\eqref{para2}. Using \eqref{para2} and \eqref{para1}, we arrive at exactly \eqref{At0}.
The new solution \eqref{mfb} can be represented by a linear scaling function $G$ as
\begin{align}
\bvf=\frac{v}{\k}G(t\k^{1/2}).
\end{align}
We show Fig.\ref{fig:I_scaling} it is consistent with numerical solutions. Comparing the scaling $t_0\sim\k^{-1/2}$ and $t_0\sim v^{-2/5}$ in phase I and noise free limit of phase II respectively, we estimate the transition between I and II to occur at $\k\sim v^{4/5}$. When $\k\gg v^{4/5}$, the quench dynamics is characterized by phase I.

\begin{figure}[htbp]
     \begin{center}
          \includegraphics[height=5cm,clip]{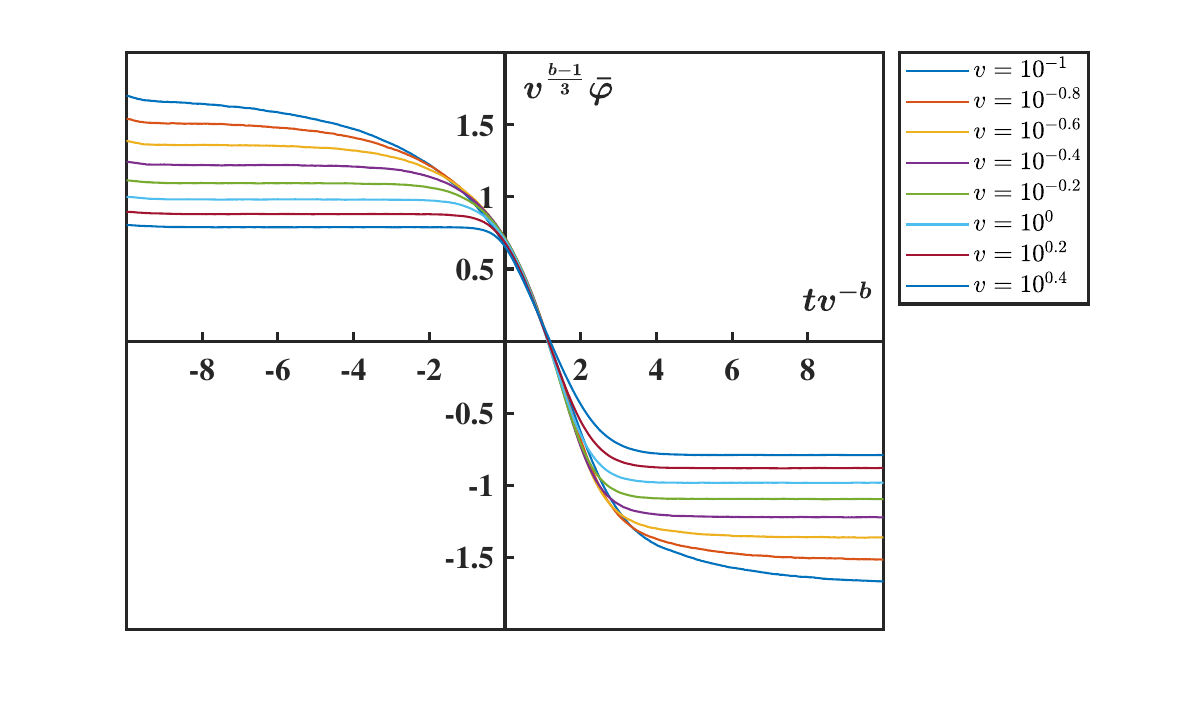}
          \includegraphics[height=5cm,clip]{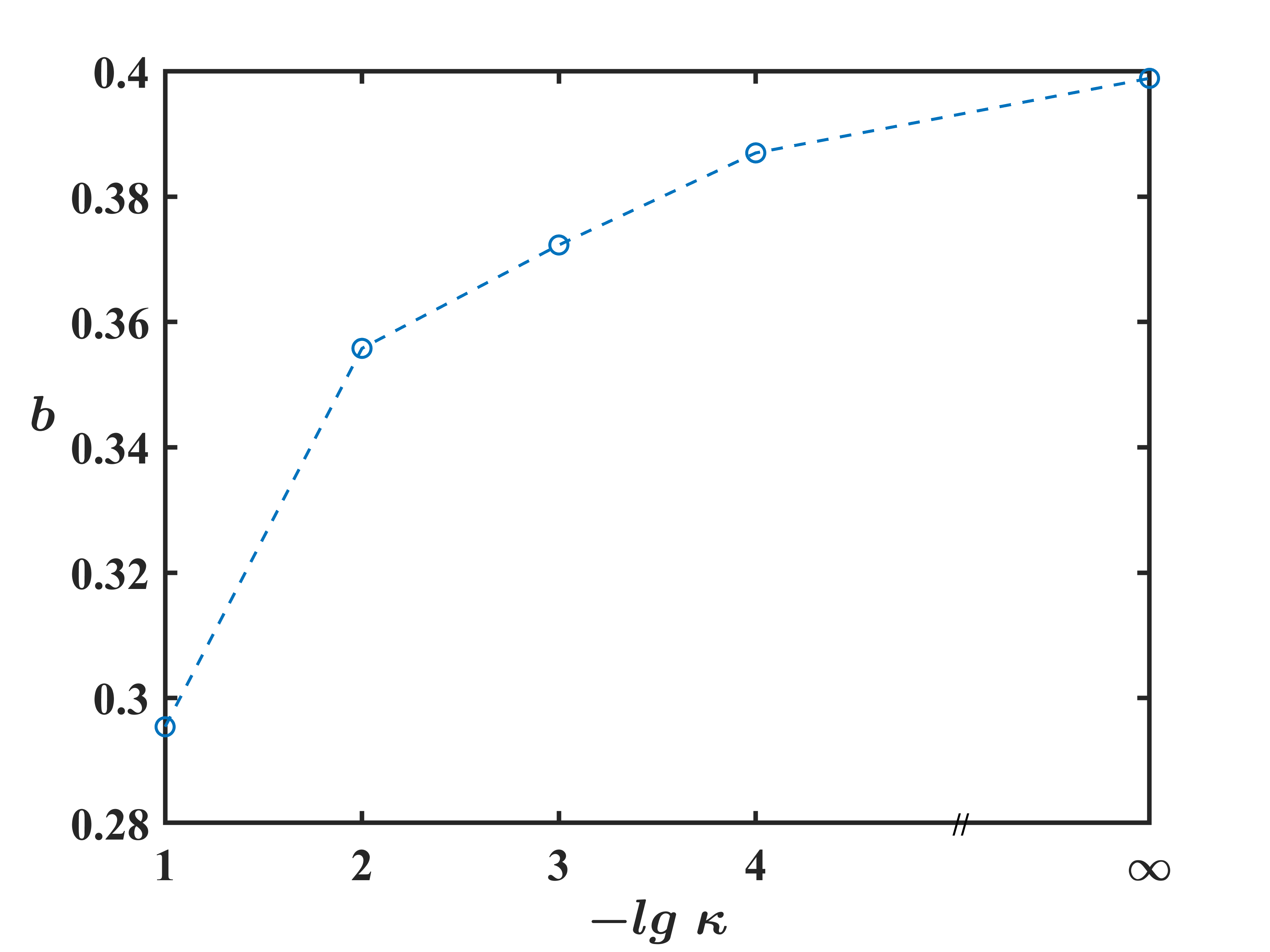}
          \caption{The upper figure shows $v^{\frac{b-1}{3}}\bvf(t)$ as a function of $tv^{b}$ for $\k=0.1$ and $b=0.29$ for intermediate values of $v$ in phase II. The lower figure shows $b$ versus $\k$. $b$ decreases with increasing $\k$ giving a shorter relaxation time.}
    \label{fig:II_scaling}
    \end{center}
\end{figure}

In the opposite limit $\k\ll v^{4/5}$ and slow quench (not very large $v$), the system is in a fluctuation modified KZ regime. In this regime, we have two possible contributions to the effective mass square. One is dynamically generated through the cubic term as $m_\eff^2\sim\bvf^2\sim J^{2/3}$, with the second relation valid at the onset of non-adiabaticity. The other is generated from fluctuation $\ce\<\D\vf^2\>$ with $\ce=3+\<\D\vf^3\>/\bvf$ (see discussion below \eqref{At0}). The mechanism is the similar to phase I but now we have $\bvf^2\gg\<\D\vf^2\>$ instead. 
The fluctuation induced correction to effective mass leads to a shorter relaxation time. However a simple scaling behavior is lost. Nevertheless we found numerically a modified KZ scaling form still exists with $t_\kz\sim v^{-b}$ with a $\k$-dependent exponent $b$. Since larger fluctuation leads to shorter relaxation time, we expect $b$ to decrease with $\k$ for $v\lesssim 1$\footnote{This is to ensure the system remains in phase II.}. On the other hand, since mean field approximation is valid, we can use the same reasoning for \eqref{scaling} to arrive at
\begin{align}
\bvf\sim J^{1/3}\sim(v t_\kz)^{1/3}\bvf\sim v^{(1-b)/3}.
\end{align}
In Fig.~\ref{fig:II_scaling}, we plot the scaling function for $\k=0.1$ in proper range of $v$, which shows a good scaling behavior. We also plot $b$ versus $\k$, confirming our expectation that as $\k$ increases, the exponent $b$ decreases. The analysis above indicates that source induced contribution to $m_\eff^2$ decreases with increasing $\k$, while fluctuation induced contribution obviously increases. It is natural to expect that at some critical strength of fluctuation, the latter contribution dominates the former, marking the disappearance of KZ scaling. This is confirmed by numerical solutions.

\begin{figure}[htbp]
     \begin{center}
          \includegraphics[height=5cm,clip]{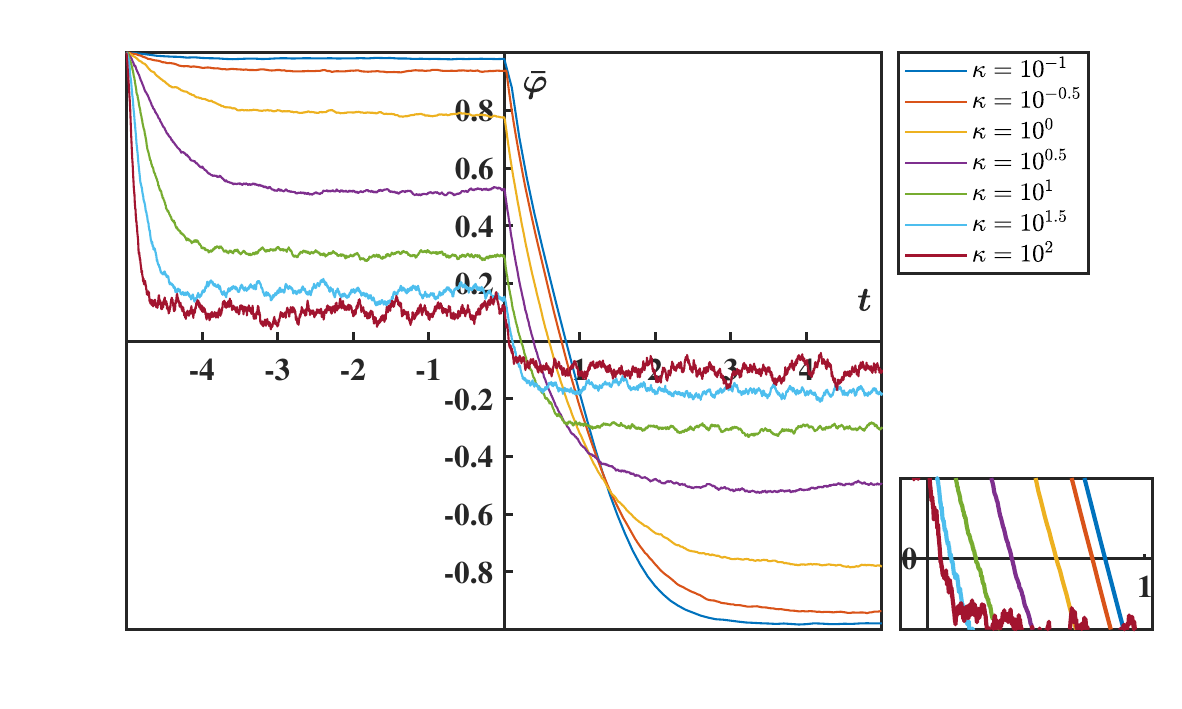}
          \caption{Time evolution of $\bvf$ for $v=10^5$ and finite $\k$ covering both phase I and phase III. The right lower panel is a zoom-in of the region where $\bvf$ crosses zero. $t_0$ at $\k=10^{-1}$ agrees well with the analytic prediction for $\k=0$ below \eqref{rapid}. 
          }
    \label{fig:III_scaling}
    \end{center}
\end{figure}

Phase III is studied numerically for large $v$ and finite $\k$. Turning on finite $\k$ enhances the relaxation, we expect a smaller $t_0$. The $\k$ dependence of $\bvf$ for a large $v$ is shown in Fig.~\ref{fig:III_scaling}, confirming that a larger $\k$ leads to a shorter $t_0$. 

Now we turn to the limits of large noise and rapid quench. We can again solve for $\bvf$ in the limit $v\to\infty$ like in the noise free case. In the presence of fluctuation, $\vf$ at $t<0$ or $J=-1$ satisfies equilibrium distribution. The equilibrium distribution for $J=-1$ is determined in the appendix. We find the noise suppresses the mean field as $\bvf\sim\k^{-1/2}$ in the large noise limit. The final mean field at $t>0$ or $J=1$ is suppressed similarly. It follows that $\bvf$ is small in the entire evolution so that \eqref{para2} and \eqref{para1} are still applicable.
It follows that the scaling solution for phase I \eqref{mfb} is also applicable. The only difference is in the numerical values of $a$ and $c$, which we determine using the equilibrium moments in appendix as $a\simeq 0.478$ and $ac\simeq-0.388$. They lead to the following dynamical relaxation time
\begin{align}\label{t0kappa2}
t_0\simeq 0.96\k^{-1/2}.
\end{align}
We remark that in order to justify the approximation $J\simeq vt_0$ in deriving \eqref{mfb}, we need to have $vt_0\ll1$, or equivalently $v^{-1}\gg \k^{-1/2}$. This has the simple interpretation that the variation of the source is much slower than the relaxation of the system. In a sense, this regime is physically similar to phase I with the fluctuation induced relaxation determining $t_0$.

In the opposite limit $v\gg \k^{1/2}$ but still $\k\gg1$, we can obtain an analytic solution in the limit $v\to\infty$ following the same logic as the noise free case. By using \eqref{para2}, \eqref{para1} and the equilibrium initial condition $\bvf(t=0)=d\k^{-1/2}$ with $d=\frac{2\sqrt{2}\G(\frac{3}{4})}{\G(\frac{1}{4})}$ from \eqref{limits}\footnote{The reason for this initial condition can be seen from Fig.~\ref{fig:III_scaling}. Time evolution of $\bvf$ quickly settles to the equilibrium value after a quick equilibration process.}:
\begin{align}\label{largevk}
\bvf=\frac{e^{-C\k^{1/2}t}(1+dC-e^{C\k^{1/2}t})}{C\k^{1/2}},
\end{align}
where $C=a(3+c)\simeq 1.046$. In arriving at \eqref{largevk}, we have dropped the subleading $\bvf^3$ term in solving \eqref{mean} because $\bvf\sim \k^{-1/2}\ll 1$. \eqref{largevk} gives the following crossing time
\begin{align}\label{t0kappa}
t_0=\frac{\ln(1+dC)}{C\k^{1/2}}\simeq 0.66\k^{-1/2}.
\end{align}
We have confirmed that \eqref{t0kappa} becomes accurate already at $\k\gtrsim 10^{0.5}$ with numerical solutions. We stress that although \eqref{t0kappa} is qualitatively the same as \eqref{t0kappa2}, the mechanism is slightly different: in this case the variation of the source is much quicker than the relaxation of the system. The dynamics is similar to the noise free limit of phase III, with the difference being in the initial and final vevs of $\bvf$ in the evolution, which are suppressed by noise as $\bvf_{\text{i/f}}\sim\k^{-1/2}$. It is the suppressed range of variation of $\bvf$ that determines the scaling \eqref{t0kappa}.


We can summarize our results of the dynamical relaxation time $t_0$ in different phases in terms of inverse mass square. In the noise free limit of slow quench, the inverse mass square comes from the cubic term as $m_\eff^2\sim\bvf^2\sim J^{2/3}$, with the vev induced by the source $J$ at the onset of non-adiabaticity $t_\kz$. The inverse mass square can also arise entirely from fluctuations, such as in phase I. Phase II can be viewed as a fluctuation modified KZ scaling, with the effective mass square having contributions from both the source and the fluctuations. Phase III is an exception in that the variation of the source is much quicker than the relaxation of system so that $t_0$ is determined by the range of variation of $\bvf$, which is also dependent on fluctuations.

\section{\bf Discussions}\label{sec_finite}

We have made two major simplifications in our analysis: we have ignored the mass term by assuming the system is close to the critical point; we have also used the truncation $\frac{\int d^3x\f(t,x)^3}{V}=\vf(t)^3$. Let us try to estimate the effect of including the mass term and going beyond the truncation based on the simple picture obtained in the previous section.

For the mass term, we consider $m^2>0$\footnote{Negative mass square would lead to condensation.}, which gives an additional contribution to effective mass square. Let us compare $m^2$ with the fluctuation induced effective mass square $\sim\k^{1/2}$. Using that $m^2\sim \x^{-2}$ and $\k=\frac{2T}{V}\sim L^{-3}T$, we find
\begin{align}
\frac{m^2}{\k^{1/2}}\sim\frac{L^{3/2}T^{1/2}}{\x^2}.
\end{align}
When $L^{3/2}T^{1/2}\gg\x^2$, the mass induced relaxation dominates over the fluctuation induced relaxation. When $L^{3/2}T^{1/2}\ll\x^2$, the finite size effect becomes significant and the fluctuations play a major role in relaxation of the system.

Next we turn to the cubic term $\frac{\int d^3x\f(t,x)^3}{V}$. In a finite volume, we can decompose $\f(t,x)$ into discrete Fourier modes $\vf_n(t)$
\begin{align}
\f(t,x)=\sum_{\vec{n}}\vf_{\vec{n}}(t)e^{i\vec{n}\frac{2\p}{L}\cdot{\vec{x}}}.
\end{align}
$\vec{n}$ is a three dimensional vector with arbitrary integer components. The cubic term can then be expressed using these Fourier modes as
\begin{align}\label{cubic_rep}
\frac{\int d^3x\f(t,x)^3}{V}=\sum_{\vec{n}_1,\vec{n}_2}\vf_{\vec{n}_1}(t)\vf_{\vec{n}_2}(t)\vf_{-\vec{n}_1-\vec{n}_2}(t).
\end{align}
Keeping only the term with $\vec{n}_1=\vec{n}_2=0$ corresponds to the truncation we have used. Note that \eqref{cubic_rep} contains a mass term for $\bvf$ of the form $3\sum_{\vec{n}}\<\vf_{\vec{n}}\vf_{-\vec{n}}\>\bvf$, which also includes fluctuation of the $\vec{n}=0$ (zero momentum) mode. 
To quantify the size of the mass term from all momentum modes, we apply $\int d^3xe^{-i\vec{n}\frac{2\pi}{L}\cdot\vec{x}}/V$ to \eqref{eom_f} to obtain an analog of \eqref{eom_vf} for $n\ne0$
\begin{align}\label{eom_vfn}
&\frac{d\vf_{\vec{n}}(t)}{dt}+n^2\(\frac{2\p}{L}\)^2\vf_{\vec{n}}(t)+\sum_{\vec{n}_1,\vec{n}_2}\vf_{\vec{n}_1}(t)\vf_{\vec{n}_2}(t)\vf_{\vec{n}-\vec{n}_1-\vec{n}_2}(t)\nonumber\\
&=\x_{\vec{n}}(t),\\
&\text{and}\nonumber\\
&\<\x_{\vec{n}}(t)\>=0,\quad \<\x_{\vec{n}}(t)\x_{-\vec{n}}(t')\>=\k\d(t-t'),\nonumber
\end{align}
from Fourier transform of \eqref{white}. The strength of the noise is independent of $\vec{n}$ following from property of white noise. Crucially, in the absence of noise, only the $\vec{n}=0$ mode $\vf_{0}\equiv\vf$ is excited by the spatially homogeneous source $J$, the truncation is justified. The noise can excite modes with nonvanishing $\vec{n}$. The form of noise indicates that the excitations carry no net momentum, thus the lowest nontrivial fluctuations for $n\ne0$ are $\<\vf_{\vec{n}}(t)\vf_{-\vec{n}}(t)\>$. We estimate their magnitude by the fluctuation-dissipation theorem
\begin{align}\label{vfn_fluc}
\<\vf_{\vec{n}}(t)\vf_{-\vec{n}}(t)\>=\frac{\k\t_n}{2},
\end{align}
with $\t_n$ being the relaxation time for $\vf_{\vec{n}}$. Based on the picture we have obtained, $\t_n$ is given by the inverse effective mass from \eqref{eom_vfn}
\begin{align}\label{meff}
\t_n=m_\eff^{-2}=\frac{1}{n^2\(\frac{2\p}{L}\)^2+3\sum_{\vec{m}}\<\vf_{\vec{m}}\vf_{-\vec{m}}\>},
\end{align}
which differs from that of $\bvf$ by the finite momentum contribution. We can sum \eqref{vfn_fluc} over $n$ and use \eqref{meff} to obtain the following equation
\begin{align}
M^2=\sum_{\vec{n}}\frac{\k}{n^2\(\frac{2\p}{L}\)^2+3M^2},
\end{align}
with $M^2=\sum_{\vec{m}}\<\vf_{\vec{m}}\vf_{-\vec{m}}\>$. The summation over $\vec{n}$ is ultra-violet (UV) divergent. Note that $n\frac{2\p}{L}$ is nothing but the discrete momentum. For large volume, we can approximate the sum by momentum integral as
\begin{align}
M^2=V\int^\L\frac{d^3k}{(2\p)^3}\frac{\k}{k^2+M^2}.
\end{align}
Clearly the momentum integral is to be cutoff at an ultra-violet scale $\L$ for the effective action \eqref{Gamma}. Since $M\ll\L$, we may ignore $M^2$ in the denominator to obtain the following scaling
\begin{align}\label{Meff}
M^2\sim V\k\L\sim \L T,
\end{align}
where we have used the scaling $\k\sim TV^{-1}$. \eqref{Meff} gives the contribution to effective mass square from the fluctuations of finite momentum modes. Note that the cutoff scale and temperature are always larger than the momentum scale set by the correlation length $\L\gg\x^{-1}$, $T\gg\x^{-1}$. By comparing $M^2$ with $m^2\sim\x^{-2}$, we find the relaxation from fluctuations of finite momentum modes is always more effective than the mass term. In this case, we can indeed neglect the mass term.

It is not necessary to compare $M^2$ with $\k^{1/2}$ because the former contains fluctuations of all momentum modes, which already includes the latter from the fluctuation of zero momentum mode only. In fact, if we simply took $n=0$ and drop the summation, which corresponds to keeping the zero momentum mode only, we would obtain $M^2\sim\k^{1/2}$. This is nothing but the scaling of effective mass square we obtained for phase I in the previous section. The summation significantly enhances the fluctuations, leading to a different scaling in \eqref{Meff}.

One may ask how does the phase diagram change when fluctuation from all momentum modes are included. Recall that the phase diagram obtained in the previous section can be qualitatively understood as from an interplay of three time scales: $v^{-1}$ variation of source, $\k^{-1/2}$ inverse effective mass square and $t_\kz$ from the breaking down of adiabaticity. The analysis above suggests we should replace the effective mass square from fluctuations of zero momentum mode only by that from fluctuations of all momentum modes. It is tempting to speculate that the phase diagram would remain qualitatively the same. The effect of the enhanced fluctuations is to lead to further shrinkage of phase II, as well as phase III. Possible nontrivial interplay between the source and the fluctuations is left out in the analysis, which deserves further studies.

\section{\bf Summary}\label{sec_summary}

We have studied effect of thermal fluctuations in a sourced quench and mapped out a phase diagram of the quench dynamics. By tracing the dynamics of the zero momentum mode driven by a spatially homogeneous source, ignoring fluctuations of finite momentum modes, we have identified three phases with a dynamical relaxation time as order parameter. Phase I occurs for large fluctuations, in which the relaxation time is always set by a fluctuation induced effective thermal mass square. Phase II occurs for small fluctuation and slow quench rate. The dynamics is qualitatively similar to the noise free limit of KZ scaling, but with a modified exponent. Small fluctuation and rapid quench rate leads to phase III, in which the relaxation of system happens well behind time variation of the source, giving a finite relaxation time in the limit of very rapid quench rate.

We have also considered the fluctuations of finite momentum modes. We have estimated their contribution to effective mass square for the zero momentum mode, finding a significant enhancement. We speculate based on the picture of relaxation dynamics near the critical point that qualitative features of the phase diagram remain the same. However, the enhanced fluctuations may lead to shrinkage of phase II and phase III.

We end by making connections to critical dynamics in heavy ion collisions. Despite the model used in this study is of mean field type while the QCD critical point is of Wilson-Fisher type \cite{Tsypin:1994nh,Tsypin:1996zw}, the picture obtained in this study may shed some light on the relaxation dynamics near the critical point. If we interpret the source field as a time-dependent mass for the order parameter field, the role of the source field is similar to temperature and baryon chemical potential of the evolving quark-gluon plasma, which determine the quark mass as source to the chiral condensate field. Our results suggest fluctuations can be enhanced when contribution from all momentum modes are included. The effect of fluctuation may significantly narrow the regime of KZ scaling as found in \cite{Mukherjee:2016kyu} based on truncation to the dynamics to zero momentum mode \cite{Mukherjee:2015swa}.

\acknowledgments

SL is grateful to Chao-Hong Lee and Yi Yin for insightful discussions and Yanyan Bu and Mitsutoshi Fujita for collaboration on related works. This work is in part supported by NSFC under Grant Nos 12075328, 11735007 and 11675274.


\appendix

\section{Equilibrium distribution}

We derive equilibrium properties of the system for $J=-1$ (corresponding to $v\to\infty$ and $t<0$). The case for $J=1$ can be obtained similarly. The stochastic equation \eqref{eom_vf} is equivalent to the following Fokker-Planck equation
\begin{align}\label{FP}
\pd_tP(\vf,t)=-\pd_\vf\[(\vf^3-1)P(\vf,t)+\frac{\k}{2}\pd_\vf P(\vf,t)\].
\end{align}
It adopts the following equilibrium distribution
\begin{align}\label{dist}
P(\vf)\propto e^{-\frac{\k}{2}\(\frac{\vf^4}{4}-\vf\)}.
\end{align}
With \eqref{dist}, it is easy to work out moments of $\vf$:
\begin{align}\label{I_def}
I_n\equiv\<\vf^n\>=\int d\vf P(\vf)\vf^n.
\end{align}
The overall normalization of $P(\vf)$ is fixed by the total probability being unity $I_0=1$. Analytic expressions of $I_n$ for general $\k$ are available in terms of hypergeometric functions.
%
We only list the large $\k$ limit needed for us
\begin{align}\label{limits}
&I_1(\k\to\infty)=\frac{2\sqrt{2}\G(\frac{3}{4})}{\G(\frac{1}{4})}\k^{-1/2},\quad I_2(\k\to\infty)=\frac{2\p}{\G(\frac{1}{4})^2}\k^{1/2}.\nonumber\\
&I_3(\k)=1.
\end{align}
Interestingly, in the large noise limit \eqref{limits} indicates that $\bvf$ is suppressed. It follows that the reasoning above \eqref{para2} is also valid for equilibrium state so that the parametrizations \eqref{para2} and \eqref{para1} remain applicable.
To determine the parameters $a$ and $c$ for equilibrium distribution, we use the following relations
\begin{align}
&I_2=I_1^2+\<\D\vf^2\>,\nonumber\\
&I_3=I_1^3+3I_1\<\D\vf^2\>+\<\D\vf^3\>.
\end{align}
Noting that in the equations of above, $I_1^2$ and $I_1^3$ are negligible and using \eqref{limits}, we obtain
\begin{align}\label{a}
a=\frac{2\p}{\G(1/4)^2}\simeq 0.478,
\end{align}
and
\begin{align}\label{ac}
ac=\(1-\frac{12\sqrt{2}\p\G(3/4)}{\G(3/4)^3}\)/\(\frac{2\sqrt{2}\G(3/4)}{\G(1/4)}\)\simeq-0.388.
\end{align}

\bibliography{KZ.bib}

\end{document}